\documentstyle[mprocl]{article}

\input{psfig}

\begin{document}

\title{
PION ELECTROPRODUCTION ON THE NUCLEON AND THE GENERALIZED GDH
SUM RULE}

\author{L. TIATOR, D. DRECHSEL}

\address{Institut f\"{u}r Kernphysik, Johannes Gutenberg-Universit\"{a}t,
J.~J.~Becher-Weg 45,\\
D-55099 Mainz, Germany\\
E-mail: tiator@kph.uni-mainz.de}

\author{S.S. KAMALOV}

\address{Laboratory of Theoretical Physics, JINR Dubna, 141980 Moscow region,
Russia\\E-mail: kamalov@thsun1.jinr.ru}

\maketitle

\abstracts{Results of the recently developed unitary isobar model
(MAID) are presented for the spin asymmetries, structure functions
and relevant sum rules in the resonance region. The model
describes the presently available data for single-pion photo- and
electroproduction in the resonance region below $W_{\rm cm}= 2$
GeV very well. It is based on Born terms and resonance
contributions, and the respective multipoles are constructed in a
gauge-invariant and unitary way for each partial wave.  The eta
production is included in a similar way, while the contribution of
more-pion and higher channels is modeled by comparison with the
total cross sections and simple phenomenological assumptions. Our
evaluation of the energy-weighted integrals is in good agreement
for the proton but shows big discrepancies for the neutron. }

\section{Introduction}\label{sec:intro}
The spin structure of the nucleon in the resonance region is of
particular interest to understand the rapid transition from resonance
dominated coherent processes to incoherent processes of deep inelastic
scattering (DIS) off the constituents. By scattering polarized lepton
beams off polarized targets, it has become possible to determine the
spin structure functions $g_1$ and $g_2$. The results of the first
experiments at CERN~\cite{Ash89} and SLAC~\cite{Bau83} sparked
considerable interest in the community, because the first moment of
$g_1$, $\Gamma_1=\int^1_0g_1(x)dx$, was found to be substantially
smaller than expected from the quark model, in particular from the
Ellis-Jaffe sum rule~\cite{Ell74}.

Here we present the results of the recently developed unitary
isobar model (MAID)~\cite{Dre98a} for the spin asymmetries,
structure functions and relevant sum rules in the resonance
region. This model describes the presently available data for
single-pion photo- and electroproduction up to a total $cm$ energy
$W_{\rm max}= 2$ GeV and for $Q^2\le$ 4 (GeV/c)$^2$. It is based
on effective Lagrangians for Born terms and vector meson exchange
(background) and resonance contributions modeled by Breit-Wigner
functions. All major resonances below $W=1700$ MeV are included.
The respective multipoles are constructed in a gauge-invariant and
unitary way for each partial wave.  The eta production is included
in a similar way~\cite{Kno95}, while the contribution of more-pion
and higher channels is modeled by comparison with the total cross
sections and simple phenomenological assumptions.

\section{Formalism}\label{sec:form}

The differential cross section for exclusive electroproduction of
mesons from polarized targets using polarized electrons, e.g.
$\vec p(\vec e,e'\pi^0)p$ can be parametrized in terms of 18
response functions \cite{Dre92}, a total of 36 is possible if in
addition also the recoil polarization is observed. Due to the
azimuthal symmetry most of them vanish by integration over the
angle $\phi$ and only 5 total cross sections remain. The
differential cross section for the electron is then given by
\begin{equation}
\frac{d\sigma}{d\Omega\ dE'} = \Gamma\sigma (\nu,Q^2)\,,
\label{eq1}
\end{equation}
\begin{equation}
\sigma =\sigma_T+\epsilon\sigma_L+
P_y\sqrt{2\epsilon(1+\epsilon)}\
         \sigma_{LT}+
         hP_x\sqrt{2\epsilon(1-\epsilon)}\
         \sigma_{LT'}+hP_z\sqrt{1-\epsilon^2}\sigma_{TT'}\, ,
\label{eq2}
\end{equation}
where $\Gamma$ is the flux of the
virtual photon field and the $\sigma_i,$ $i=L$, $T$, $LT$, $LT'$,
$TT'$, are functions of the $lab$ energy of the virtual photon
$\nu$ and the squared four-momentum transferred $Q^2$. These
response functions can be separated by varying the transverse
polarization $\epsilon$ of the virtual photon as well as the
polarizations of the electron ($h$) and proton ($P_z$ parallel,
$P_x$ perpendicular to the virtual photon, in the scattering plane
and $P_y$ perpendicular to the scattering plane). In particular,
$\sigma_{T}$ and $\sigma_{TT'}$ can be expressed in terms of the
total cross sections for excitation of hadronic states with spin
projections $3/2$ and $1/2$:
$\sigma_{T}=(\sigma_{3/2}+\sigma_{1/2})/2$ and
$\sigma_{TT'}=(\sigma_{3/2}-\sigma_{1/2})/2.$

Here we use Hand's notation with the equivalent photon $cm$ energy
$K=(W^2-m^2)/(2W)$ for the virtual photon flux. Correspondingly,
the phase space factors of the cross sections are given by $q/K$,
where $q$ is the pion momentum in the $cm$.

In inclusive electron scattering  $\vec e+\vec N\rightarrow X$,
only 4 cross sections $\sigma_T$, $\sigma_L$, $\sigma_{LT'}$ and
$\sigma_{TT'}$ appear, the fifth cross section, $\sigma_{LT}$,
vanishes due to unitarity when all open channels are summed up.
The individual channels, however, give finite contributions.

The Gerasimov-Drell-Hearn (GDH) sum rule is only derived for real
photons. It is based on unitarity and low-energy theorems and the
assumption of the convergence of an unsubtracted dispersion
relation,
\begin{eqnarray}
I_{GDH} &=& \frac{m^2}{8\pi^2\alpha}\int_{\nu_0}^{\infty}
            \left (\sigma_{1/2}-\sigma_{3/2}
            \right )\ \frac{d\nu}{\nu}\
            = -\frac{\kappa^2}{4} \,.
\label{eq3}
\end{eqnarray}
This sum rule is often presented without the leading factor in
front of the integral, the numerical conversion is
$8\pi^2\alpha/m^2=254.8\mu b$.

It can be generalized in various ways. Three forms often used in
the literature are summing up only contributions from
$\sigma_{TT'}$ with no longitudinal terms,
\begin{eqnarray}
I_{GDH}^{(a)}(Q^2)
&=&\frac{m^2}{8\pi^2\alpha}\int_{\nu_0}^{\infty}
           (1-x)
           \left (\sigma_{1/2}-\sigma_{3/2}
           \right )\ \frac{d\nu}{\nu}\ \,,
\label{eq4}
\\
I_{GDH}^{(b)}(Q^2)
&=&\frac{m^2}{8\pi^2\alpha}\int_{\nu_0}^{\infty}
           \frac{K}{\mid\vec{k}_{\gamma}\mid}
           \left (\sigma_{1/2}-\sigma_{3/2}
           \right )\ \frac{d\nu}{\nu}\ \,,
\label{eq5}
\\
I_{GDH}^{(c)}(Q^2)
&=&\frac{m^2}{8\pi^2\alpha}\int_{\nu_0}^{\infty}
           \left (\sigma_{1/2}-\sigma_{3/2}
           \right )\ \frac{d\nu}{\nu}\ \,.
\label{eq6}
\end{eqnarray}
The factor $K/\mid\vec{k}_{\gamma}\mid$ can also be expressed as
$(1-x)/\sqrt{1+\gamma^2}$.
The relations between the $\sigma_i$ and the quark structure
functions $g_1$ and $g_2$ can be read off the following equations,
which define further possible generalizations of the GDH
integral~\cite{Ger65} and the Burkhardt-Cottingham (BC) sum
rule~\cite{Bur70}, which in addition also include
longitudinal-transverse interference terms,
\begin{eqnarray}
I_1(Q^2) &=& \frac{2m^2}{Q^2}\int_{0}^{x_0}g_1(x,Q^2)\ dx
\nonumber
\\
&=&\frac{m^2}{8\pi^2\alpha}\int_{\nu_0}^{\infty}
           \frac{1-x} {1+\gamma^2}
           \left (\sigma_{1/2}-\sigma_{3/2}
           -2\gamma\,\sigma_{LT'}\right )\ \frac{d\nu}{\nu}\ \,,
\label{eq7}
\\
I_2(Q^2) &=& \frac{2m^2}{Q^2}\int_{0}^{x_0}g_2(x,Q^2)\ dx \nonumber
\\
&=&  \frac{m^2}{8\pi^2\alpha}\int_{\nu_0}^{\infty}
           \frac{1-x} {1+\gamma^2}
           \left (\sigma_{3/2}-\sigma_{1/2}
           -\frac{2}{\gamma}\,\sigma_{LT'}\right )\ \frac{d\nu}{\nu} \,,
\label{eq8}
\\
I_3(Q^2) &=& \frac{2m^2}{Q^2}\int_{0}^{x_0}(g_1(x,Q^2)+g_2(x,Q^2))\ dx \nonumber
\\
&=&  -\frac{m^2}{4\pi^2\alpha}\int_{\nu_0}^{\infty}
           \frac{1-x}{Q}\,\sigma_{LT'}\ d\nu\ = I_1+I_2 \,,
\label{eq9}
\end{eqnarray}
where $\gamma=Q/\nu$ and $x=Q^2/2m\nu$ the Bjorken scaling
variable, with $x_0$ ($\nu_0$) referring to the inelastic
threshold of one-pion production. Since $\sigma_{LT'}={\cal
O}(Q)$, the real photon limit of the integral $I_1$  is given by
the GDH sum rule $I_1(0)=I_{GDH}(0)=-\kappa_N^2/4,$ with
$\kappa_N$ the anomalous magnetic moment of the nucleon. At large
$Q^2$ the structure functions should depend only on $x,$ i.e.
$I_1\rightarrow 2m\Gamma_1/Q^2$ with $\Gamma_1=\int
g_1(x){\rm d}x=$ const. In the case of the proton, all experiments
for $Q^2> 1$GeV$^2$ yield $\Gamma_1>0.$ Therefore, a strong
variation of $I_1(Q^2)$ with a zero-crossing at $Q^2<1$ GeV$^2$ is
required in order to reconcile the GDH sum rule with the
measurements in the DIS region. The $I_2$ integral of
Eq.~(\ref{eq8}) is constrained by the BC sum rule, which requires
that the inelastic contribution for $0 < x <x_0 $ equals the
negative value of the elastic contribution, i.e.
\begin{equation}
I_2(Q^2) = \frac{2m^2}{Q^2}\int_{0}^{x_0}g_2(x,Q^2)\ dx
=\frac{1}{4}\frac{G_M(Q^2)-G_E(Q^2)}{1+Q^2/4m^2}\,G_M(Q^2)\,,
\label{eq10}
\end{equation}
where $G_M$ and $G_E$ are the magnetic and electric Sachs form
factors respectively. At large $Q^2$ the integral vanishes as
$Q^{-10}$, while at the real photon limit
$I_2(0)=\kappa_N^2/4+e_N\kappa_N/4$, the two terms on the right
hand side corresponding to the contributions of $\sigma_{TT'}$ and
$\sigma_{LT'}$ respectively. Finally, Eq. (\ref{eq9}) defines an
integral $I_3(Q^2)$ as the sum of $I_1(Q^2)$ and $I_2(Q^2)$ and is
given by the unweighted integral over the longitudinal-transverse
interference cross section $\sigma_{LT'}$. At the real photon
point this integral is given by the GDH and BC sum rules,
$I_3(0)=e_N\kappa_N/4$. In particular this vanishes for the
neutron target.

\section{Unitary Isobar Model}\label{sec:uim}
Our calculation for the response functions $\sigma_i$ is based on
the Unitary Isobar Model (UIM) for one-pion photo- and
electroproduction of Ref.~\cite{Dre98a}, accessible in the
internet as the MAID program. The model is constructed with
effective Lagrangians for Born terms, vector meson exchange in the
$t$ channel (background), and the dominant resonances up to the
third resonance region are modeled using Breit-Wigner functions
with energy-dependent widths. For each partial wave the multipoles
satisfy gauge invariance and unitarity. As in any realistic model
a special effort is needed to describe the $s$-channel multipoles
$S_{11}$ and $S_{31}$. Even close at threshold these multipoles
pick up sizeable imaginary parts that cannot be explained by
nucleon resonances. In fact the $S_{11}(1535)$, $S_{11}(1650)$ and
the $S_{31}(1620)$ play only a minor role for the complex phase of
the $E_{0+}$ multipoles even at higher energies. The main effect
arises from pion rescattering. This we can take into account by
$K$-matrix unitarization. Furthermore we introduce a gauge
invariant contact term proportional to the anomalous magnetic
moment of the nucleon $\kappa_N$,
\begin{equation}
j_\kappa^\mu = \frac{ieg}{2m}\kappa_N
F(q_0^2)\frac{\sigma^{\mu\nu}k_\nu}{2m}\gamma_5 \,.
\label{eq11}
\end{equation}
The form factor $F(q_0^2)$, $q_0$ being the asymptotic pion
momentum, vanishes at threshold, consistent with chiral symmetry,
but gives rise to a cancellation of unphysically high momentum
components in the Born terms at high energies.

Due to unitarity each partial wave has to fulfill Watson's
theorem,
\begin{eqnarray}
t_{\gamma,\pi}^\alpha &=& t_{\gamma,\pi}^\alpha(background) +
t_{\gamma,\pi}^\alpha(resonances)\\ \nonumber &=& \pm \mid
t_{\gamma,\pi}^\alpha \mid e^{i\delta_{\pi N}^\alpha} \, .
\label{eq12}
\end{eqnarray}
In an isobar model this condition has to be constructed
explicitly. In Maid98 the background is real (except for the
S-waves) and a phase is added to the resonance. In Maid2000 both
background and resonance contributions are unitarized separately
for all partial waves up to $l=3$ in the following way
\begin{equation}
   t_{\gamma,\pi}^\alpha = t_{\gamma,\pi}^\alpha(Born+\omega,\rho)
   (1\,+\,i\,t_{\pi N}^I) +
   t_{\gamma,\pi}^\alpha(resonances)e^{i\Psi^\alpha}\, .
\label{eq13}
\end{equation}

\begin{figure}[ht]
\centerline{
\psfig{file=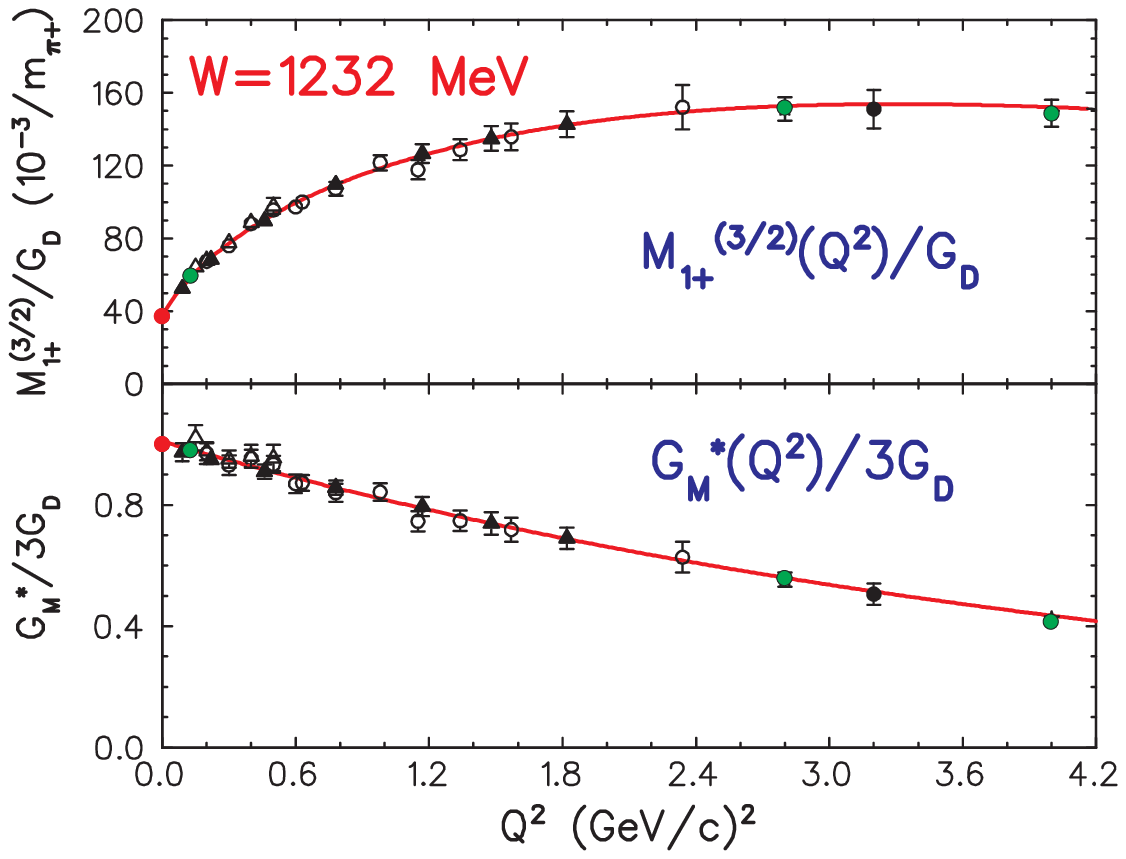,height=4.5cm,angle=0,silent=}
\hspace{0.1cm}
\psfig{file=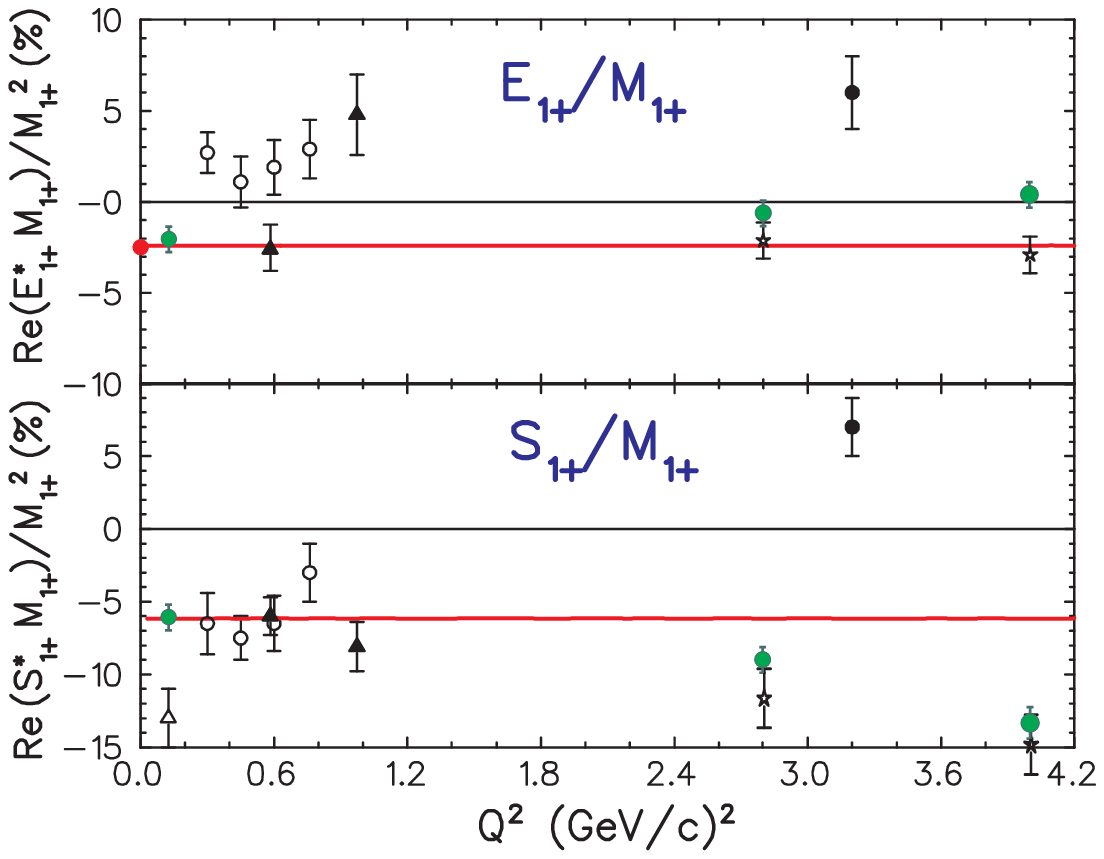,height=4.5cm,angle=0,silent=} }
\caption{\label{fig1} $M_{1+}^{(3/2)}$ multipole and magnetic
transition form factor $G_M^*$ divided by the nucleon dipole form
factor $G_D$ and E/M and S/M ratios of the $\Delta$ at the resonance
position $W=1232$ MeV. The gray circles at $Q^2=0.13$ GeV$^2$ show our
analysis of the
Bates~\protect\cite{mertz99} data and at $Q^2=2.8$ and $4.0$ GeV$^2$
for the JLab~\protect\cite{frol99} data. For further details see
Ref.~\protect\cite{Dre98a}.}
\end{figure}
In Fig. \ref{fig1} we show the $M_{1+}^{(3/2)}$ multipole and the
$M1$ transition form factor of the $\Delta$ resonance on the left
and the $R_{EM}=E_{1+}^{3/2}/M_{1+}^{3/2}$ and
$R_{SM}=S_{1+}^{3/2}/M_{1+}^{3/2}$ ratios on the right side. The
experimental data agree very well with our empirical fit for
$G_M^*$. For the small ratios the
experimental information is not yet reliable enough to justify
anything else than a constant value. In our model we use
$R_{EM}=-2.2\%$ and $R_{SM}=-6.5\%$. Recent analyses of
experimental data on $p(e,e'\pi^0)p$ from JLab~\cite{frol99} show
a trend to $E/M$ ratios very close to zero and increasing negative
values for $S/M$ at large $Q^2$.

\begin{figure}[ht]
\centerline{ \psfig{file=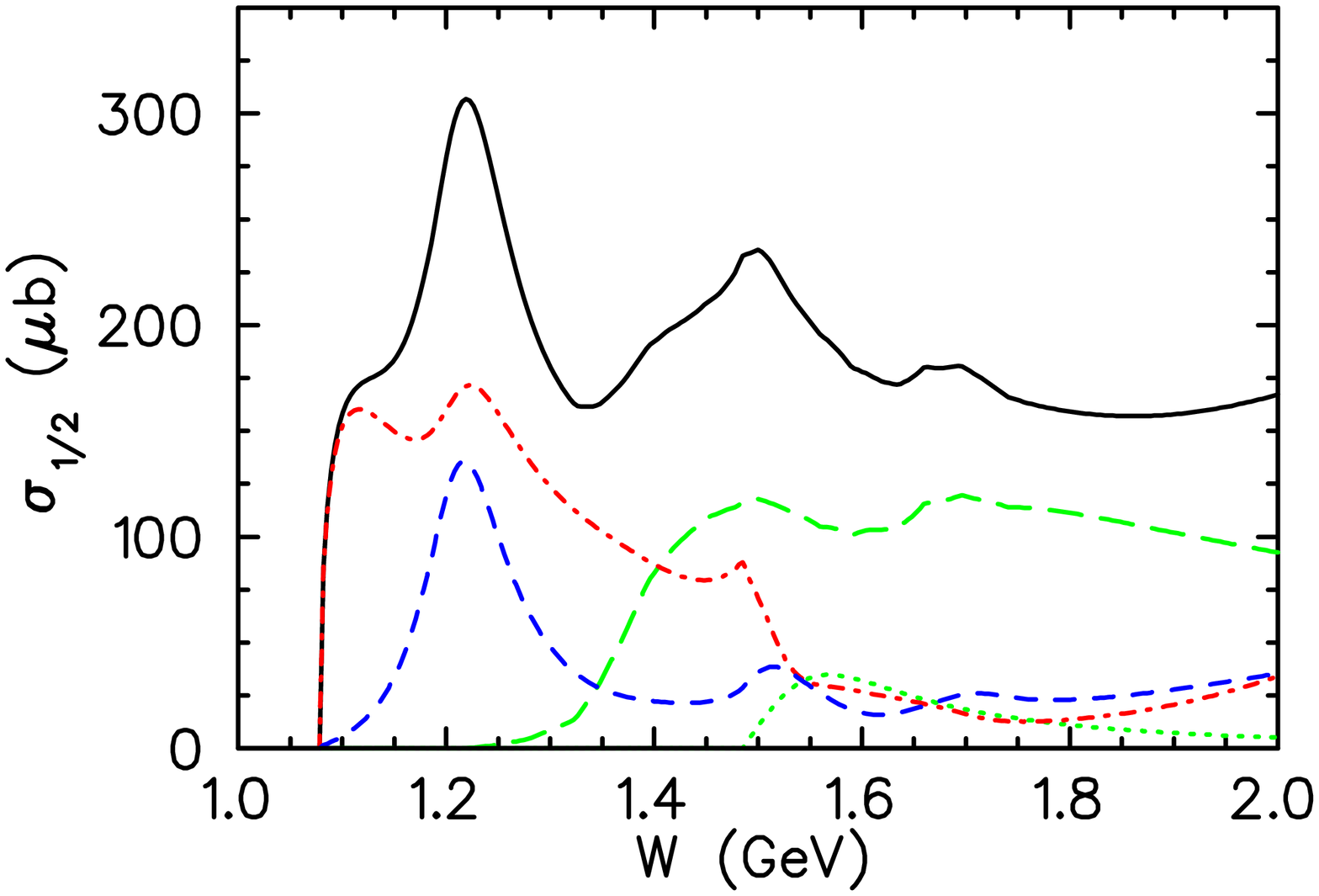,height=4.15cm,angle=90,silent=}
\hspace{0.1cm}
\psfig{file=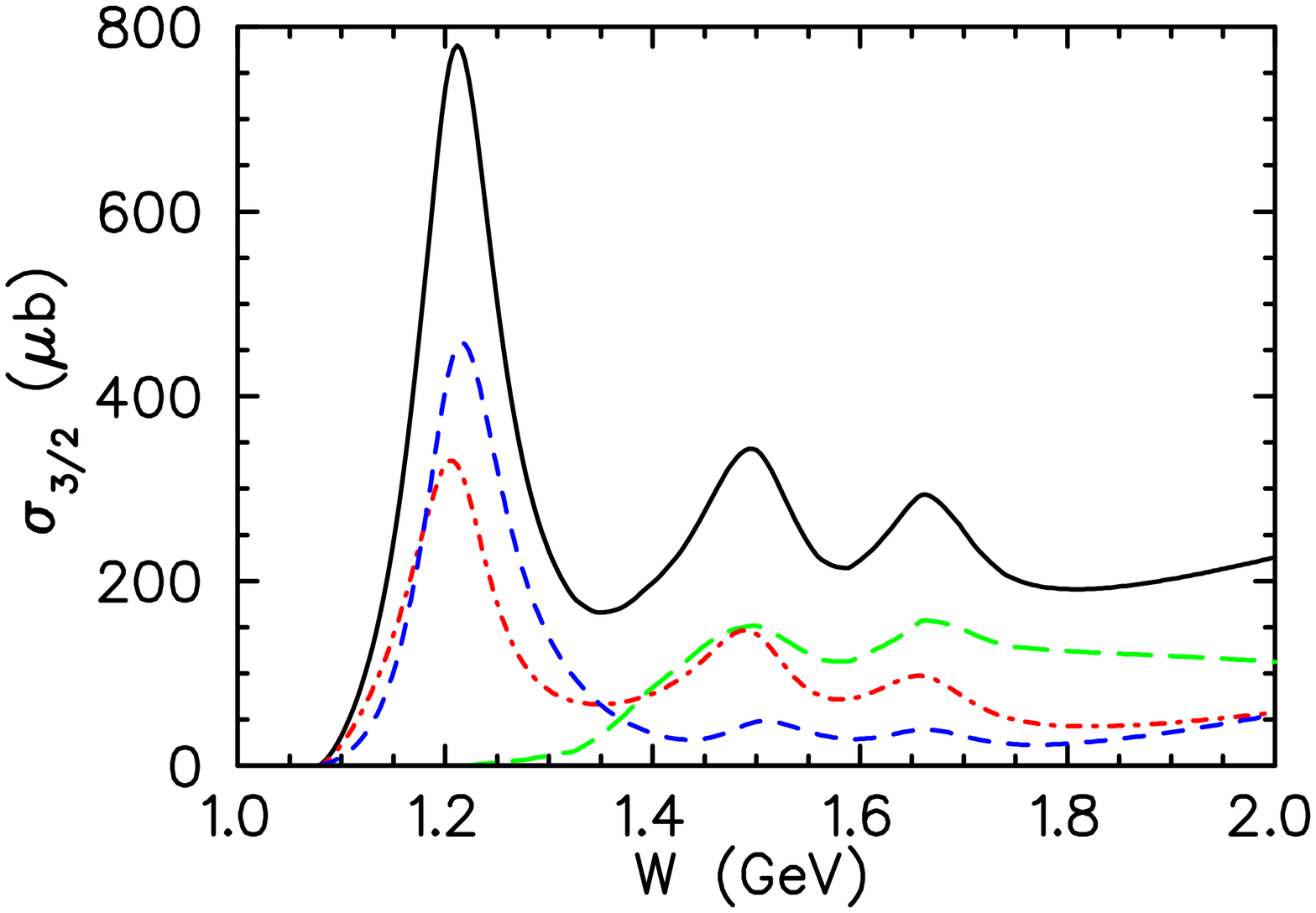,height=4.15cm,angle=90,silent=} }
\vspace{0.5cm} \centerline{
\psfig{file=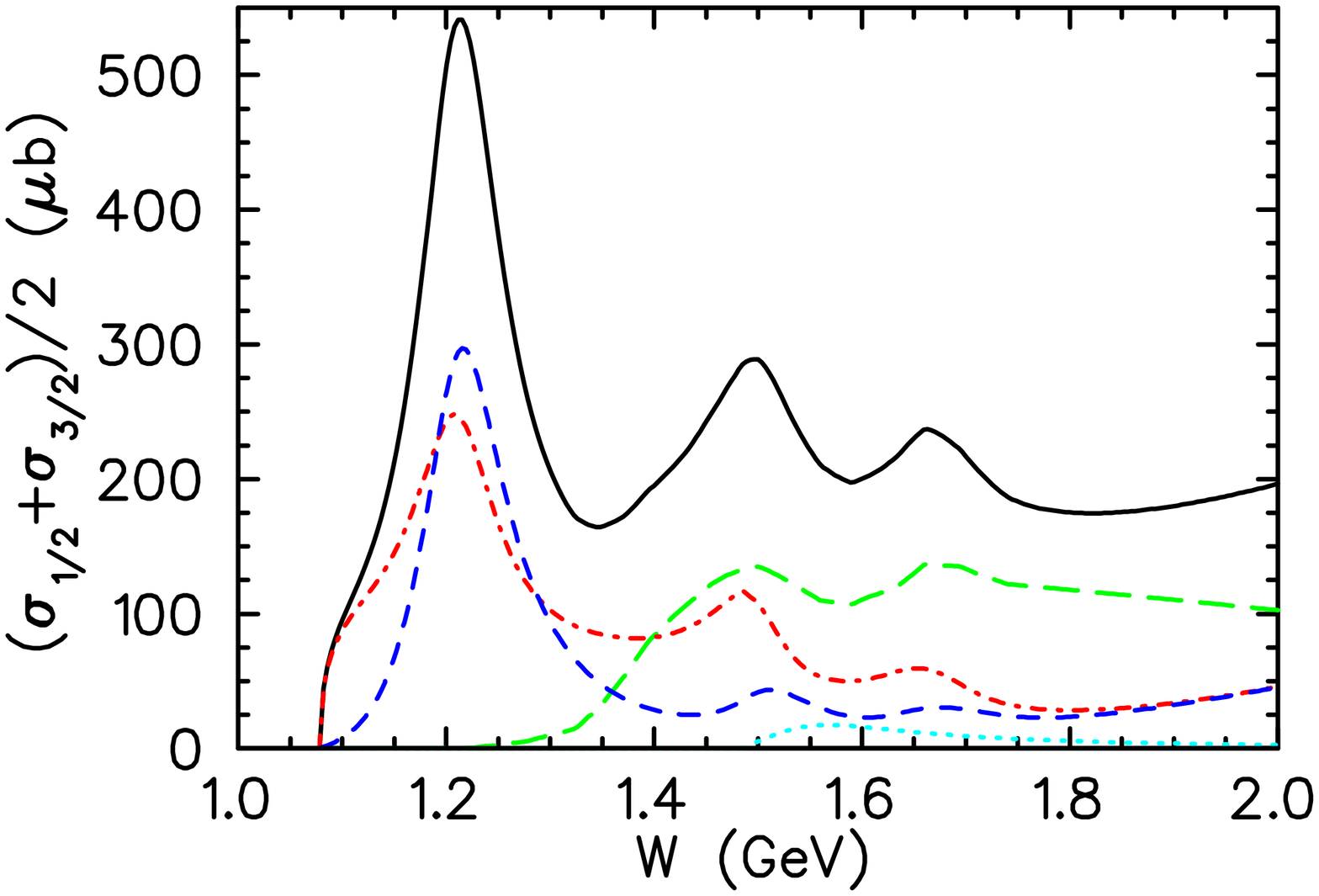,height=4.15cm,angle=90,silent=}
\hspace{0.1cm}
\psfig{file=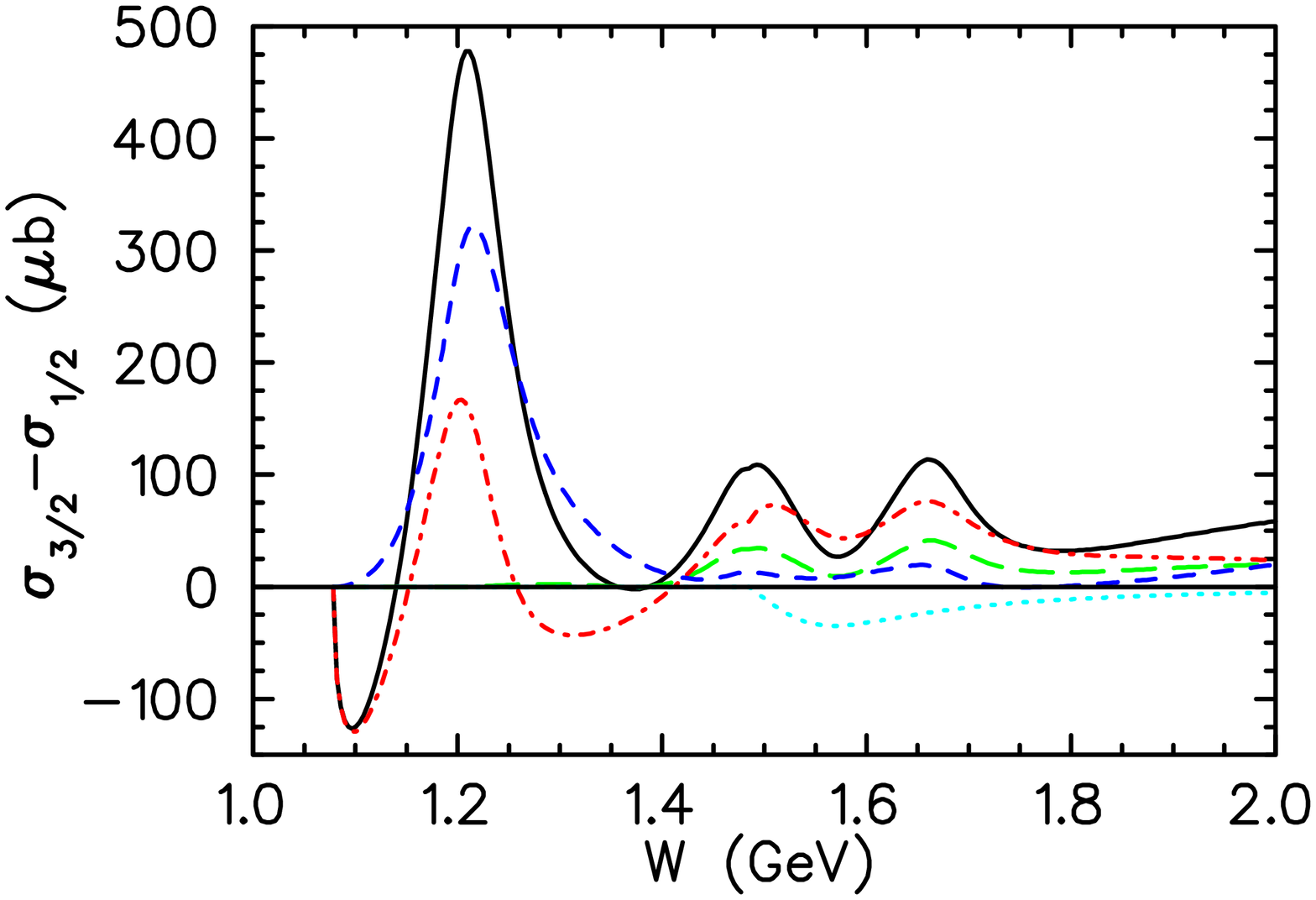,height=4.15cm,angle=90,silent=} }
\caption{\label{fig2}
  Helicity dependent cross sections for photoproduction on the proton target.
  The dashed, dash-dotted, long dashed and dotted lines show our
  calculations for $\pi^0$, $\pi^+$, two-pion and the $\eta$ cross
  sections, respectively. The solid curves give the total
  photoabsorption and contain the sum of all.}
\end{figure}

The UIM is able to describe the single-pion electroproduction
channel quite well. However, at higher energies the contributions
from other channels become increasingly important. In the
structure functions $\sigma_{T}$ and $\sigma_{TT'}$ we account for
the $\eta$ and the multi-pion production contributions extracting
the necessary information from the existing data for the total
cross section~\cite{Dre98b}. In Fig.~\ref{fig2} we show the
individual channels for the total helicity dependent cross
sections $\sigma_{1/2}$, $\sigma_{3/2}$, $\sigma_T$ and
$\Delta\sigma=2\sigma_{TT'}$ at $Q^2=0$.
Due to the non-regularized Born terms
in the $1\pi$ channels the cross sections start to rise again at
energies $W>1.8$ GeV. However, because of the energy weighting,
the effect is negligible for the integrals.

\section{Integrals}\label{sec:int}
\subsection{Results for Real Photons}
In Tab.~\ref{tab1} we show our results for the GDH integral and
the forward spin polarizability over the lab energy range of
200-450 MeV together with the latest Mainz
data~\cite{ahre00}. In comparison with the results of the
dispersion theoretical partial wave analysis HDT~\cite{HDT}
and the SAID solution SM99K~\cite{SAID} our MAID results agree
very well with the experiment. The additional information on the
individual channels, however, offers interesting insights in the
different calculations, especially the $\pi^+$ result of SAID
for $\gamma_0$ is a factor 3 larger than MAID and about a
factor 2 above the data. The reason is the enhanced sensitivity
of background contributions in the $\pi^+$ channel, especially
the S-wave near threshold.

Tab.~\ref{tab2} shows the GDH integral over
the full energy range up to $W_{max}=2$~GeV. The preliminary
experimental result is obtained only from measurements at Mainz
and covers the energy range from 200 to 800 MeV photon $lab$
energy. However, our theoretical calculations indicate a very
close cancellation between the low energy contribution from
threshold up to 200 MeV (30 $\mu b$) and of energies above 800 MeV
(-34$\mu b$). For our detailed comparison we included recent
calculations of reggeized $\pi\pi$ photoproduction by Holvoet and
Vanderhaeghen \cite{holv00} that include $\gamma,\pi\Delta$ Born
terms and additional $D_{13}(1520)$ excitations. This $2\pi$
contribution to the GDH integral is about twice as large as compared
to our simple phenomenological multi-pion parametrization used for
finite $Q^2$.

Our calculation that also include very recent Regge-type
calculations for two-pion photoproduction~\cite{holv00} shows a
very good agreement with the sum rule for the proton target but
also exhibits a big deviation for the neutron target. However, on
the neutron target the photoproduction information is rather
limited above the $\Delta$ region and it has to be further
investigated if the high-energy region is perhaps more important
than for the proton target. Furthermore, for both nucleon targets
high energy contributions beyond the two-pion production have to
be studied that make a very big contribution in deep inelastic
scattering at finite $Q^2 > 1 GeV^2$.

\begin{table}[htbp]
\begin{center}
\caption{\label{tab1} Contributions to the  GDH integral
$I=\int_{200}^{450}(\sigma_{1/2}-\sigma_{3/2})/\nu d\nu$ and to
the forward spin polarizability of the proton
$\gamma_0=1/4\pi^2\int_{200}^{450}(\sigma_{1/2}-\sigma_{3/2})/\nu^3
d\nu$ for photon $lab$ energies of 200-450 MeV.}
\vskip 0.3 cm
\begin{tabular}{ccccc}
\hline
 $I (\mu b)$ & Mainz exp. \cite{ahre00}  & SM99K \cite{SAID} &
 HDT \cite{HDT} & MAID \cite{Dre98a} \\ \hline
 $\pi^0 p$ & -124 $\pm$ 11 & -132 & -144 & -136 \\
 $\pi^+ n$ &  -33 $\pm$ 3  &  -55 &  -26 &  -23 \\
 total     & -157 $\pm$ 11 & -187 & -170 & -159 \\
\hline
 $\gamma_0 (10^{-4}fm^4)$ &   &  &  &  \\ \hline
 $\pi^0 p$ & -1.2 $\pm$ 0.3 & -1.34 & -1.48 & -1.40 \\
 $\pi^+ n$ & -0.23$\pm$ 0.04& -0.54 & -0.19 & -0.17 \\
 total     & -1.4 $\pm$ 0.3 & -1.88 & -1.67 & -1.57 \\
\hline
\end{tabular}
\end{center}
\end{table}

\begin{table}[htbp]
\begin{center}
\caption{\label{tab2} Contributions to the GDH integral for proton
and neutron: Sum rules $-2\pi^2\alpha\kappa_N/m^2$ (sr), Mainz
experiment \protect\cite{brag00} in the energy interval of 200-800
MeV , MAID $1\pi$ contributions, eta production
\protect\cite{Kno95}, reggeized $2\pi$ contributions (Born terms
and $D_{13}(1520)$ resonance) \protect\cite{holv00}.} \vskip 0.3cm
\begin{tabular}{ccccccccc}
\hline
 $I (\mu b)$ & sr & exp. &
 $\gamma,\pi^0$ & $\gamma,\pi^\pm$ &
 $\gamma,\eta$ & $\gamma,\pi\pi\, B$ &
 $\gamma,\pi\pi\, D_{13}$& sum \\
 \hline
 proton  & -205 & -228 $\pm$ 6 & -150 & -21 & +15 & -30 & -15 & -201 \\
 neutron & -233 &      ---     & -154 & +30 & +10 & -35 & -15 & -164 \\
\hline
\end{tabular}
\end{center}
\end{table}

\subsection{Results for Virtual Photons}
\begin{figure}[htbp]
\centerline{
\psfig{file=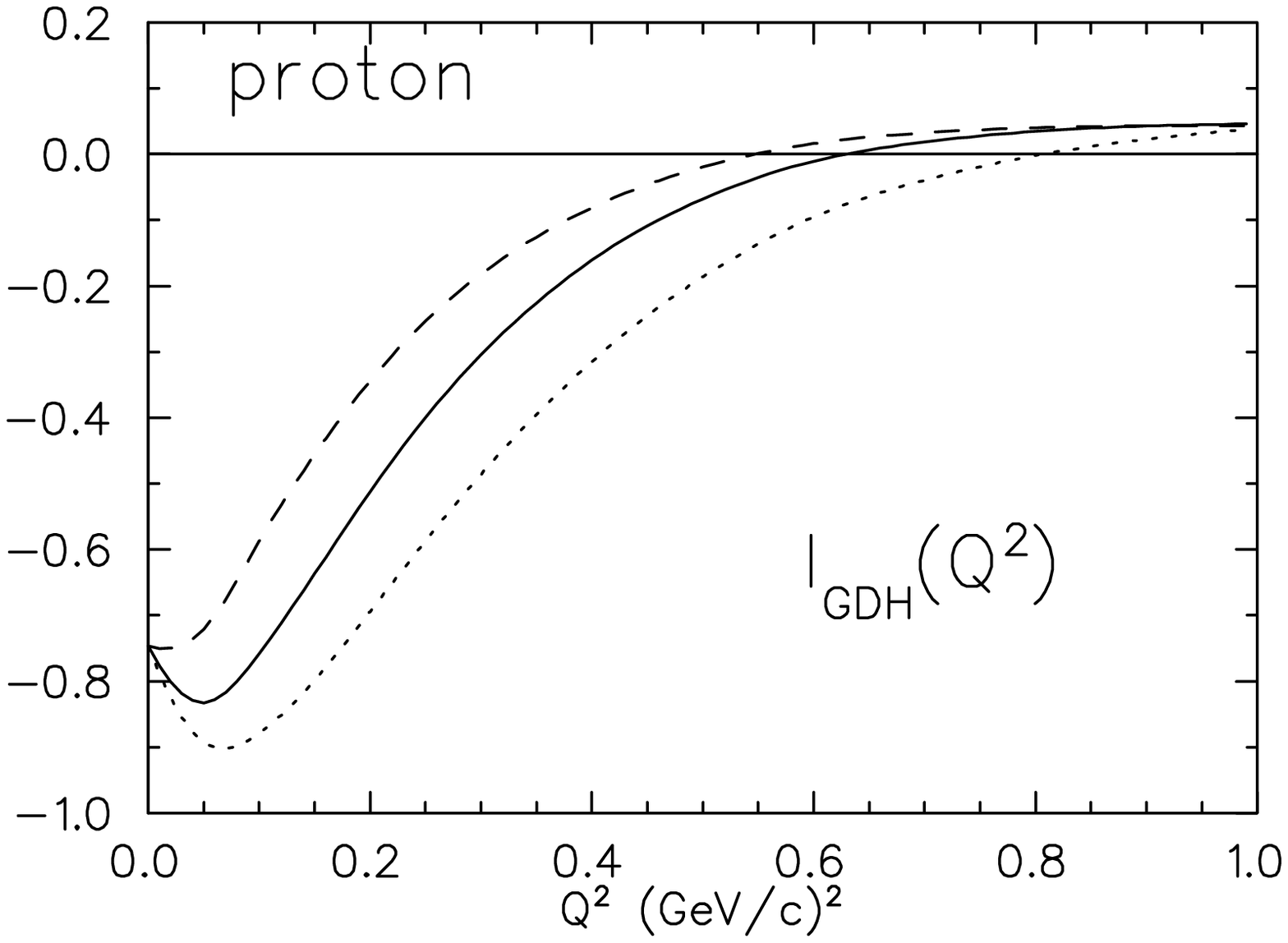,height=4.4cm,angle=90,silent=}
\hspace{0.1cm}
\psfig{file=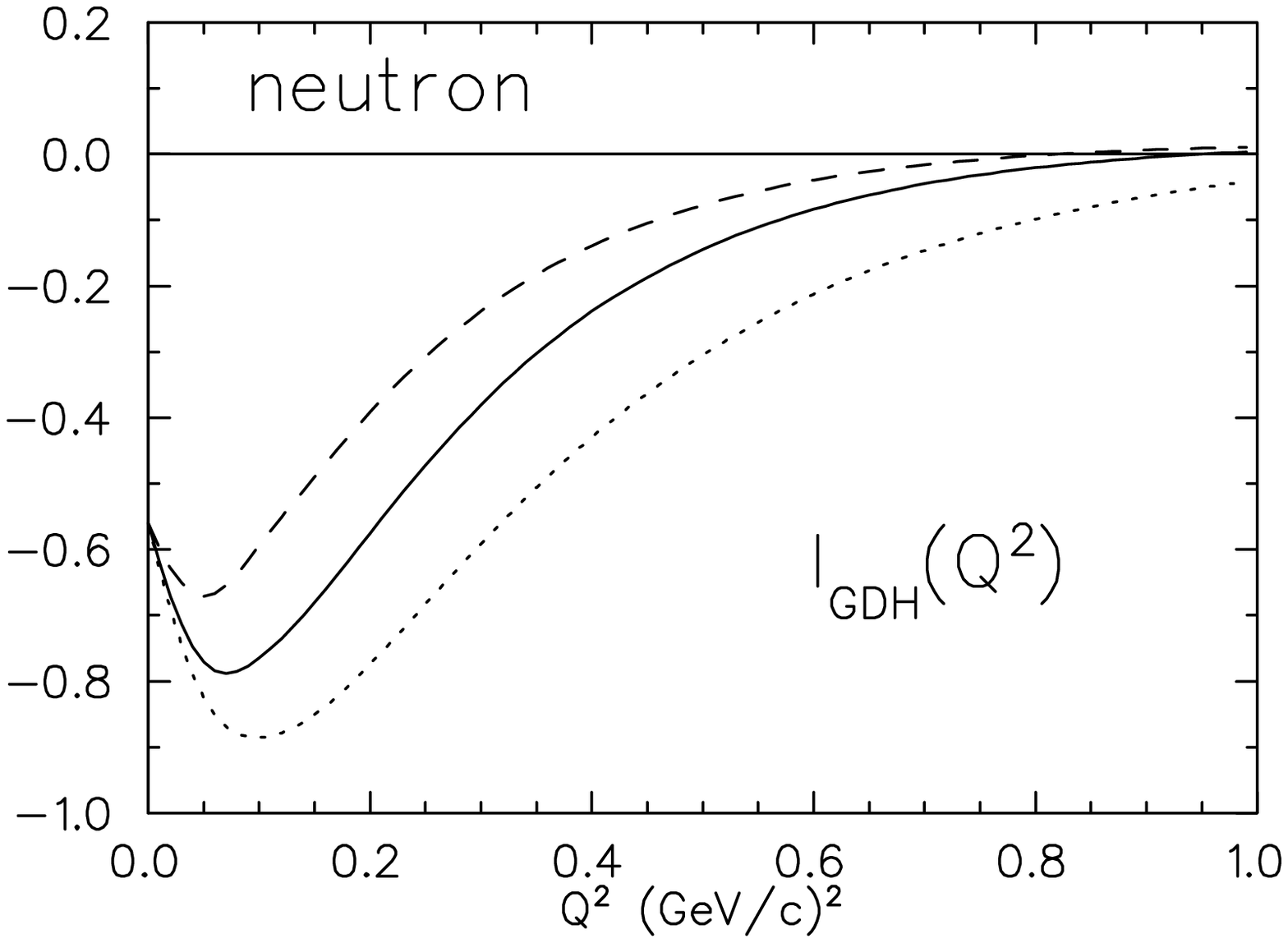,height=4.4cm,angle=90,silent=} }
\caption{\label{fig3} Generalized GDH integrals $I_{GDH}(Q^2)$ for
   3 different definitions used in the literature. The full, dashed
   and dotted lines show the integrals (a), (b), (c) in the notation
   of Eqs. \ref{eq4}, \ref{eq5}, \ref{eq6}, respectively. The integrals
   are evaluated up to $W=2$~GeV and include $1\pi + \eta +n\pi$
   contributions.}
\end{figure}

\begin{figure}[htb]
\centerline{\psfig{file=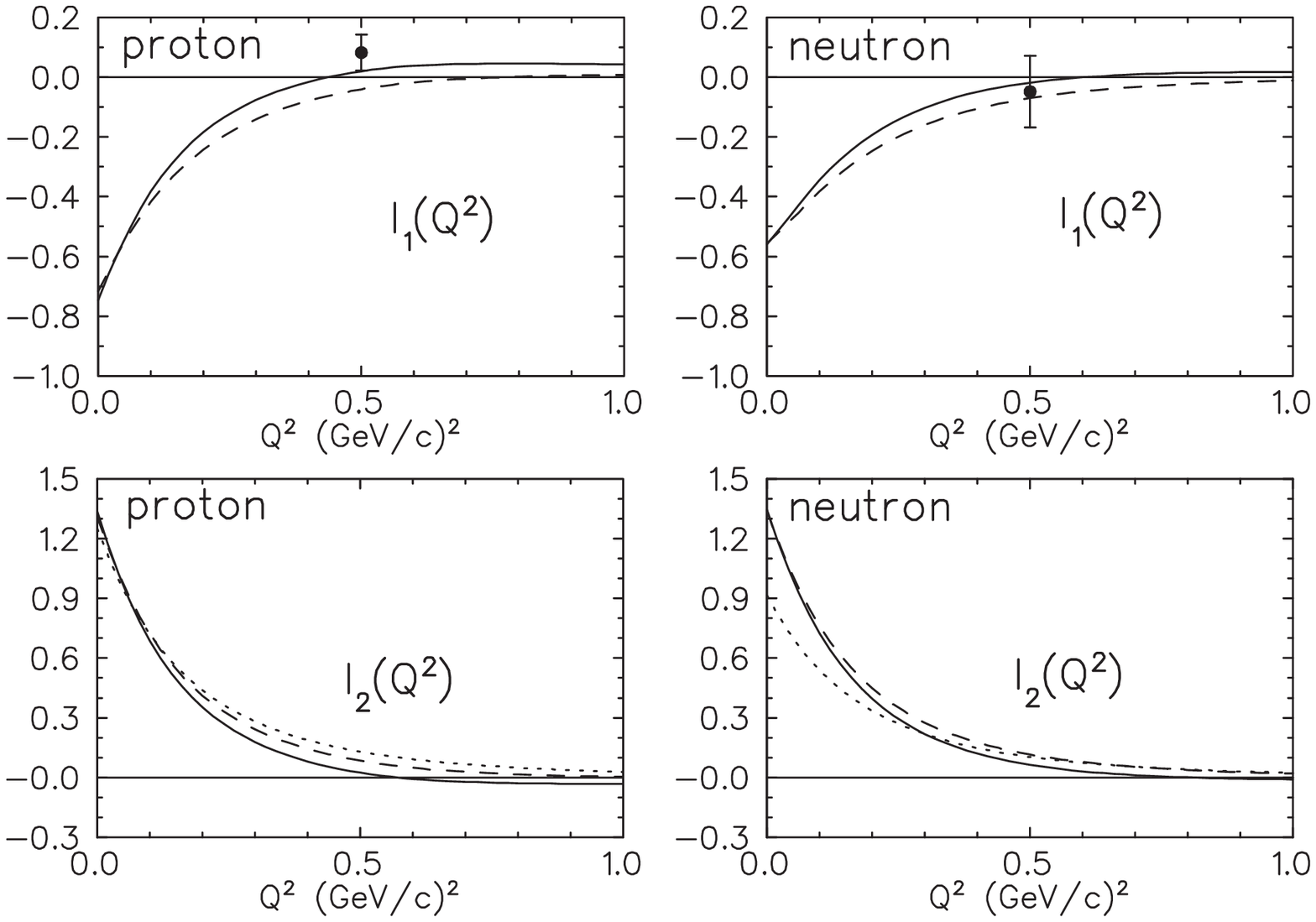,height=8.0cm,angle=0,silent=}}
\caption{\label{fig4} Generalized GDH integrals $I_{1,2}(Q^2)$ for
the proton and the neutron integrated up to
  $W_{{\rm max}}=2$ GeV. The dashed
  lines show the contributions from the $1\pi$ channel while the full
  lines include $1\pi + \eta +n\pi$.
  The dotted line for $I_2$ is the $BC$ sum rule prediction
  of~{Ref.~\protect\cite{Bur70}}.
  The data is from SLAC, Ref.\protect\cite{Abe98}.}
\end{figure}

In Fig.~\ref{fig3} and Fig.~\ref{fig4}  we give our predictions
for the integrals $I_{GDH}(Q^2)$, $I_1(Q^2)$ and $I_2(Q^2)$ in the
resonance region, i.e. integrated up to $W_{{\rm max}}=2$ GeV for
the proton and neutron targets. A comparison of the 3 different
forms, defined in Eqs. \ref{eq4}, \ref{eq5}, \ref{eq6} show
significantly different slopes at $Q^2=0$ and quite different zero
positions, where the GDH integral crosses from negative values
observed for real photons to positive values known from deep
inelastic scattering. In the case of the integral $I_1,$ our model
is able to generate the expected drastic change in the helicity
structure at low $Q^2$.
We find a zero-crossing at $Q^2=0.75$
(GeV/c)$^2$ if we include only the one-pion contribution. This
value is lowered to 0.52 (GeV/c)$^2$ and 0.45 (GeV/c)$^2$ when we
include the $\eta$ and the multi-pion contributions respectively.
The SLAC analysis of the proton yields $I_1=0.1\pm0.06$ at $Q^2=
0.5$ (GeV/c)$^2$, while our result at this point is only slightly
positive. For the neutron our calculation is fully consistent
within the SLAC analysis at $Q^2= 0.5$ (GeV/c)$^2$ in contrast to
the large discrepancy observed at $Q^2=0$, see Tab. 2.

Comparing with the generalizations of the GDH sum rule in
Fig.~\ref{fig3}, it can be seen that the slope at $Q^2=0$ and the
existence of a minimum for small $Q^2$  depends on the inclusion
of the longitudinal contributions, i.e. the minimum disappears when
$\sigma_{LT'}$ is added. Concerning the integral $I_2$,
our full result is in good agreement with the prediction of the BC
sum rule.  The deviation is within $10~\%$ and should be
attributed to contributions beyond $W_{\rm max}=2$ GeV and the
uncertainties in our calculation for $\sigma_{LT'}$. As seen in
Eq. (\ref{eq9}) the integral $I_3$ depends only on this
$\sigma_{LT'}$ contribution. From the sum rule result a value of
$e_N\kappa_N/4$ is expected at $Q^2=0$, i.e. 0.45 for the proton
and zero for the neutron target. While our value arising entirely
from the $1\pi$ channel (0.59) gets relatively close to
the sum rule result for the proton, in the neutron case this sum
rule is heavily violated  (0.78). So far it is not clear
where such a large negative contribution should arise for the
neutron target in order to cancel the $1\pi$ contribution.
Either it is due to the high-energy tail that may
converge rather slowly for the unweighted integral $I_3$, or the
multi-pion channels could contribute in such a way, while the eta
channel is very unlikely. On the other hand the convergence of the
BC sum rule cannot be given for granted. In fact Ioffe et al.
\cite{Iof84} have argued that the BC sum rule is valid only in the
scaling region, while it is violated by higher twist terms at low
$Q^2$. In any case a careful study of the multi-pion contribution
for both proton and neutron targets will be very helpful, in
particular one can expect longitudinal contributions from the
non-resonant background.

\section{Summary}

In summary, we have applied our recently developed unitary isobar model
for pion electroproduction to calculate generalized GDH integrals and
the BC sum rule for both proton and neutron targets. Our results
indicate that both the experimental analysis and the theoretical models
have to be quite accurate in order to fully describe the helicity
structure of the cross section in the resonance region.

While our results agree quite well for the GDH and BC sum rules for the
proton, we find substantial deviations for the neutron target, in
particular the sum rule $I_3(0)\equiv I_1(0)+I_2(0)=0$ is heavily
violated by the contribution from the single-pion channel which is even
larger than in the case of the proton.
Concerning the theoretical description, the treatment of the multi-pion
channels has to be improved with more refined models. On the
experimental side, the upcoming results from measurements with
real~\cite{Ahr92} and virtual photons~\cite{Bur91} hold the promise to
provide new precision data in the resonance region.

\section*{Acknowledgments}
 We would like to thank M. Vanderhaeghen for the contributions to the
 $\pi\pi$ channels and J.\ Arends for the information on the experimental data analysis.
 This work was supported by the Deutsche Forschungsgemeinschaft (SFB 443).


\end{document}